\documentclass[reprint,prx,aps,preprintnumbers,superscriptaddress,amsfonts,amssymb]{revtex4-2}                                                                                                                                                                                                                                                                                                                                                                                                                                                                                                                                                                                                                                                                                                                                                                                                                                                                                                                                                                                                                                                                                                                                                                                                                                                                                                                                 

\usepackage{graphicx}% Include figure files
\usepackage{dcolumn}% Align table columns on decimal point

\usepackage[bottom]{footmisc}

\usepackage{mathrsfs}

\usepackage{lipsum}
\usepackage[utf8]{inputenc}
\usepackage[english]{babel}
\usepackage[T1]{fontenc}
\usepackage{amsfonts}
\usepackage{amsmath}
\usepackage{amssymb}
\usepackage{graphicx}
\usepackage{caption}
\usepackage{subcaption}
\usepackage{enumitem}
\usepackage{mathtools}
\usepackage{float}
\usepackage[titletoc]{appendix}
\usepackage{multirow}
\usepackage{xcolor}
\usepackage[colorlinks=false, urlcolor=blue, pdfborder={0 0 0}]{hyperref}

\newsavebox{\bigimage}
\newcommand{\pa}{|\!+\! \alpha\rangle}

\newcommand{\beq}{\begin{equation}}
\newcommand{\eeq}{\end{equation}}
\newcommand{\beqr}{\begin{eqnarray}}
\newcommand{\eeqr}{\end{eqnarray}}
\newcommand{\ket}{\rangle}

\newcommand{\ah}{\hat{a}}
\newcommand{\ahd}{\hat{a}^\dagger}
\newcommand{\ahdsquared}{\hat{a}^{\dagger 2}}

\newcommand{\mia }{ | \mspace{-4mu}- \mspace{-3mu} \alpha \mspace{-1mu} \rangle }
\newcommand{\posa}{ | \mspace{-4mu}+ \mspace{-3mu}\alpha\mspace{-1mu} \rangle }
\newcommand{\mpa}{ | \mspace{-4mu}\pm\mspace{-3mu}\alpha \mspace{-1mu} \rangle }

\newcommand{\mib}{|\mspace{-4mu}-\mspace{-6mu}\beta \rangle}
\newcommand{\posb }{|\mspace{-4mu}+\mspace{-6mu}\beta \rangle}
\newcommand{\mpb}{|\mspace{-4mu}\pm\mspace{-4mu}\beta  \rangle}

\newcommand{\mibc}{|\mspace{-4mu}-\mspace{-6mu}\beta_c \rangle}

\newcommand{\cplusket}{\big{|}C^{+}_{\beta_t}[\phi(t)]\big{\rangle}}
\newcommand{\cplusbra}{\big{\langle} C^{+}_{\beta_{t}}[\phi(t)]\big{|}}

\newcommand{\cminusket}{\big{|}C^{-}_{\beta_t} [\phi(t)]\big{\rangle}}
\newcommand{\cminusbra}{\big{\langle} C^{-}_{\beta_t} [\phi(t)]\big{|}}

\newcommand{\Dcplusket}{\big{|} \partial_t C^{+}_{\beta_t} [\phi(t)]\big{\rangle}}
\newcommand{\Dcminusket}{\big{|} \partial_t C^{-}_{\beta_t} [\phi(t)]\big{\rangle}}
\newcommand{\Dcpmket}{\big{|} \partial_t C^{\pm}_{\beta_t} [\phi(t)]\big{\rangle}}
 
\newcommand{\cpmket}{\big{|}C^{\pm}_{\beta_t}[\phi(t)]\big{\rangle}}

\newcommand{\pcs}{P_{\textup{cs}}}

\makeatletter
\def\flushboth{%
      \let\\\@normalcr
        \@rightskip\z@skip \rightskip\@rightskip
          \leftskip\z@skip
            \parindent 1.5em\relax}
\makeatother

\begin{document}

\preprint{}

\title{Fast Gates of Detuned Cat Qubit}

\author{Arne \surname{Schlabes}}
\affiliation{Institute for Quantum Information,
RWTH Aachen University,  Aachen 52056,}
\affiliation{Peter Gr\"unberg Institute PGI-2, Forschungszentrum J\"ulich, J\"ulich 52428 , Germany}
\author{Rahul \surname{Bhowmick}}
\affiliation{Indian Institute of Science Education and Research, Kolkata}
\author{M. H. Ansari}
\affiliation{Institute for Quantum Information,
RWTH Aachen University,  Aachen 52056,}
\affiliation{Peter Gr\"unberg Institute PGI-2, Forschungszentrum J\"ulich, J\"ulich 52428 , Germany}

\date{\today}

\captionsetup[figure]{name={FIG.},labelsep=period, singlelinecheck=off}
%\captionsetup[subfigure]{position=top,singlelinecheck=off,justification=raggedright}

\begin{abstract}
Cat qubits have emerged as a promising candidate for quantum computation due to their higher error-correction thresholds and low resource overheads. In existing literature, the detuning of the two-photon drive is assumed to be zero for implementing single and multi-qubit gates. We explore a modification of the Hamiltonian for a range of detuning and demonstrate that high fidelity single qubit gates can be performed even by proper parameter matching. We also analyze the CNOT gate in presence of an approximate detuning term and explain its fidelity improvements through Shortcut to Adiabaticity corrections.
\end{abstract}

\maketitle

\section{Introduction}

Quantum processors have the potential to outperform classical computation \cite{nielsen2001quantum,shor1999polynomial,grover1997quantum,yung2014introduction,arute2019quantum}. Gate-based quantum processors encodes information in delicate many-body states made of combining quantum bits. Such states need to remain resilient against errors \cite{knill1998resilient,aharonov1997fault,Xu_2023,gottesman1998theory,preskill1998reliable,Ansari_2011,Ansari_2015}. Superconducting qubits are one of the most scalable candidates suitable for building many-body quantum states \cite{huang2020superconducting,ansari2019superconducting,xu2024lattice}. The qubits interact electromagnetically yet are prone to errors \cite{ku2020suppression,xu2021zz,wilen2021correlated}. Error correction suggest binding several physical qubits into a `logical' qubit. For example, the multi-qubit physical state $|0000\cdots\rangle$ represents the logical $|0\rangle$ \cite{shor1995scheme}, which enables detecting error and correcting it; however, for applicable computation, millions of physical qubits are required to represent a hundred logical qubits \cite{fowler2012surface}. 

Protection of quantum information against error requires further improvements in the quality of qubits and gates or coming up with new forms of qubits. A recent novel idea proposes encoding information on spatially-extended qubits, like a nonlinear cavity, namely the 'Cat' qubit, in which a cavity of microwave frequency $\omega_q$ is coupled to a set of Josephson junction-based nonlinear inductances that supply nonlinearity $\chi$ for the photons in the cavity \cite{Guillaud2022}.  The dynamics of such a qubit can be described by the Hamiltonian $H_q=\omega_q \ahd \ah +\sum_{m=3} \chi_m (\ahd + \ah )^m$,  with $\ahd$ ($\ah$) being photon creation (annihilation) operators and $\chi_m$ its $m$-th order nonlinearity. This cavity driven by a squeezing pulse of two-photon frequency $\omega_d\approx 2 \omega_q$ with Hamiltonian $H_d = \Omega \cos(\omega_d t) ( \ah^2 +  \ah^{\dagger 2})$ stabilizes single mode photons, \cite{Guillaud2022}. In a frame rotating with half of the squeezing frequency the total Hamiltonian becomes:
\begin{equation}
H=\delta \ahd \ah  - K(\chi_m) \ahdsquared  \ah^2 + \epsilon_2(\chi_m, \Omega) \left( \ahdsquared + \ah^2 \right) \label{Ham:cat}
\end{equation}
with $\delta=\omega_q- \omega_d/2+\delta'(\chi_m, \Omega)$ being the \emph{effective} harmonic frequency of photons in the driven cavity; noticing the contribution of anharmonicity $\delta'(\chi_m, \Omega)$. The nonlinearity orders result in the Kerr nonlinearity $K(\chi_m)$, similarly the squeezing drive amplitude effectively will be $\epsilon_2(\Omega,\chi_m )$.  The Kerr-Cat qubits are set to have negligible $\delta$ as reducing this parameter seems to help achieving higher fidelity in single qubit rotation \cite{Guillaud2022}. This simplifies the Hamiltonian (\ref{Ham:cat}) to $H=-K(\hat{a}^{\dagger 2}-\epsilon_2/K)(\ah^2 -\epsilon_2/K)+\epsilon_2^2/K$, which contains a pair of degenerate coherent states $\mpa $ as ground states, distant from the vacuum by $\alpha\equiv \sqrt{\epsilon_2/K}$ \citep{Guillaud2022,Ma_2021}. These states can be approximately mapped on the poles of  $Z$-axis in the Bloch sphere, producing the superpositions $|C^\pm_\alpha\rangle=\mathcal{N}^\pm_\alpha( \pa \pm \mia    ) $ on  $X$-axis,  with the normalization factor $\mathcal{N}^\pm_\alpha = 1/\sqrt{2(1\pm e^{-2|\alpha|^2})}$. Phase-flip makes stochastic $\pi$-rotations around $Z$-axis; bit-flip does similar along $X$-axis. By increasing $\alpha$ the coherent states exponentially reduce their overlap, which reduces the rate of flips \citep{Puri2019}. 

In a zero-detuned cavity $\delta=0$, $Z$ rotation takes place by further driving  Hamiltonian (\ref{Ham:cat})  by a single photon $\epsilon_z \ahd + \epsilon_z^*\ah$, which causes a relatively fast Rabi oscillation $\cos (\Omega_z t) |C^-_\alpha\rangle+ \sin(\Omega_z t)|C^+_\alpha\rangle$ with $\Omega_z=\text{Re}(4\epsilon_z \alpha)$,  \citep{Grimm2020}.  X rotation takes place in the absence of all drives. In fact by turning off the drives the Kerr nonlinearity alone evolves the coherent state $\pa$, so that the coherent state starts to dephase shortly until it phase-collapses, where phase dispersion across photon number distribution becomes $\sim \pi$. This dephasing is suppressed at the revival time $T_{rev} / 2=\pi / 2K$ when the Cat-state is reproduced.  This shows that the Kerr nonlinearity serves as a discrete $\pi/2$-rotation generator along $X$ axis \citep{Kirchmair2013}.  Increasing  Kerr nonlinearity $K$ speeds up the gate and this strengthening can take place by coupling cavity to a Superconducting Nonlinear Asymmetric Inductive eLement (SNAIL) \citep{Frattini2017}.  One consequence of discreteness of $X$ rotation is that other angles require a combination of $X$ and $Z$ rotations which prolongs gate time and reduces its fidelity \citep{{Puri2017}, {Puri2020}}.  

Here we introduce qubit rotations in the presence of finite detuning $\delta$ and show that it may not even be necessary to free up cat qubits from detuning. A detuned Cat (dCat) qubit of selected parameters can make fast and high fidelity $Z$-rotations. In contrast to detuning-free cat qubits in which only discrete X rotation are possible, the $X$ gate on dCat qubits show single-shot rotation to all angles.  This helps to prepare superpositions in a single-shot gate instead of using multiple X and Z rotations as in detuning-free cat qubits. The two gates together make universal single-qubit gates for dCat qubits. This helps to relax the delicate constraint of zeroing detuning. Such a constraint in multi-qubit processors adds further complexity in scaling up the number of qubits and relaxing the constraint can be experimentally advantageous, as a difficult tune-up sequence is required so that Stark shifts and detuning cancel out \cite{Grimm2020}.

\section{Model}  
\begin{figure}[t]
	\begin{subfigure}[t]{0.25\textwidth}
		\caption*{(a)}\vspace*{-1em}
		\includegraphics[width=1.1 \textwidth]{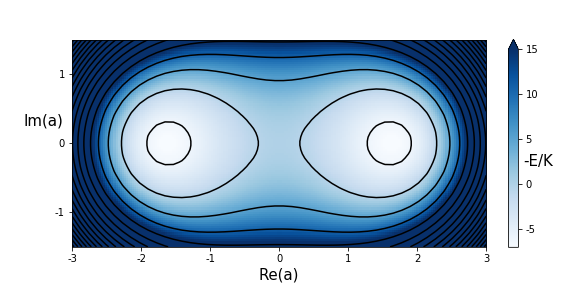}
	\end{subfigure}%
	%\hfill
	\begin{subfigure}[t]{0.25\textwidth}
		\caption*{(c)}\vspace*{-1em}
		\includegraphics[width=1.1\textwidth]{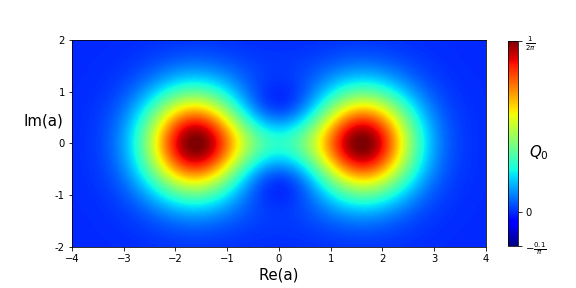}
	\end{subfigure}%
	\vspace*{-0.5em}
		\begin{subfigure}[t]{0.25\textwidth}
		\caption*{(b)}\vspace*{-1em}
		\includegraphics[width=1.1\textwidth]{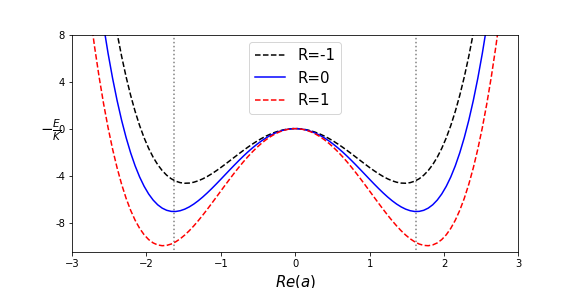}
	\end{subfigure}%
	%\hfill
	\begin{subfigure}[t]{0.25\textwidth}
		\caption*{(d)}\vspace*{-1em}
		\includegraphics[width=1.1\textwidth]{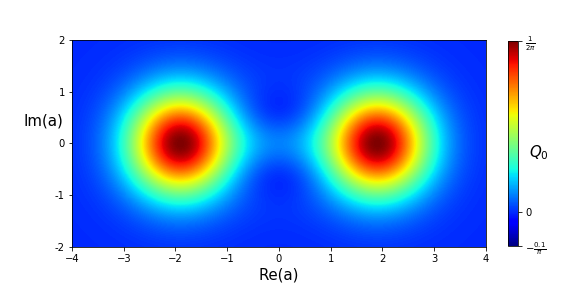}
	\end{subfigure}%
\caption{\flushboth (a) Contour of the energy of the Hamiltonian (\ref{Ham:cat}) with parameters $K/2\pi=6.7\,\text{MHz}$ and $\alpha=1.63$. (b) Energy along the real axis for zero, negative and positive detuning. Negative values compress the potential inwards, while positive values stretch it outwards. The minima are positioned at $\pm\beta$ in the complex plane and depend on $R$. (c) and (d) Husimi $Q$-function $Q_0$ of the states $(\posb  +\mib )/\sqrt{2}$ for the same parameters as in (a) with no detuning (c) and a detuning of $R=2$ (d). The state in (c) with no detuning coincides with $(\pa +\mia     )/\sqrt{2}$, since $\alpha=\beta$ for $R=0$. As the potential and its minima are stretched outwards, we shift our basis states in the same way. This means that the states in (d) are slightly further apart than in (c), stemming from the fact that the potential is equally stretched outwards. If the states were kept at $\pm\alpha$, they would become excited states inside the potential wells, which would result in them oscillating inside these wells and collecting infidelities over time.}\label{fig:1}
\end{figure}

The dynamics of a dCat qubit follow from the Hamiltonian (\ref{Ham:cat}). Without loss of generality, we can consider a finite and fixed Kerr nonlinearity in the cavity and use it as a scale to normalize the squeezing drive and detuning terms, i.e. 
\begin{equation}\label{scaledHamil}
    H/K = R \ahd \ah - (\ahdsquared  -\alpha^2)({\ah}^{2}  -\alpha^2),
\end{equation} 
with $R\equiv \delta/K$.  In the absence of a squeezing drive, i.e. $\epsilon_2=0$, the ground state of the Hamiltonian is $|0\rangle$. When the squeezing drive is switched on without detuning (R=0), the two eigenstates of the Hamiltonian are the coherent states $\mpa $, where $\alpha = \sqrt{\epsilon_2/K}$. These coherent states are degenerate as they both correspond to zero energy.  The squeezing drive $\epsilon_2$ produces an energy barrier in between the two states $\mpa $ and makes them distinct, which can be used for quantum computing, see Fig. (\ref{fig:1}b).

In the presence of detuning, i.e. $R\neq 0$, the coherent states $\mpa $ are no longer eigenstates of the Hamiltonian. The most suitable method of choosing the basis states is to maintain coherent states as the closest approximation to eigenstates that only slightly deviates from the ground state. This, however, cannot be a reliable approximation in strong detuning regime as such detuning scrambles coherent states. Therefore we limit the detuning term to be weaker than its Kerr nonlinearity, i.e. $R \lesssim 1$, which limits driving dCat qubits to be within the \emph{detuned regime}. We introduce the coherent states $\mpb  $ to best describe the lowest energy levels of the Hamiltonian (\ref{scaledHamil}) with $\beta=\sqrt{\alpha^2+R/2}$. This choice of states localizes the two eigenstates $\mpb $ in the vicinity of $\pm \alpha$, see Fig. \ref{fig:1}(c) and (d).  

The coherent $\mpb $ states are separated states by the energy barrier $E_{\textup{gap}}=4K\bar{n}$, which is insensitive to the detuning $\delta$, see Fig. (\ref{fig:1}b).  Any transition between these two states requires overcoming this barrier and the energy required for this grows with the increase of $K$ and the average number of photons in the cavity $\bar{n}$. In the intermediate detuned regime with $K \leq \delta$, the two states $\mpb $  are well-separated states and making transition between them requires an excitation to states higher or equal in energy compared to $E_{\textup{gap}}$ or a tunneling through the barrier, which is also exponentially suppressed in $E_{\textup{gap}}$.

An important remark is that the real eigenstates of Hamiltonian (\ref{scaledHamil}) can be slightly different from $\mpb $ states, showing rather different energy barrier that could be $\delta$-dependent. This later on shows the slight variation in the gate performance.

 \section{X gate for dCats}
The states $\mpb  $ evolve under the Hamiltonian (\ref{scaledHamil}) by rotating along $X$ axis; however, as mentioned earlier, these states are different from being the eigenstates. This mismatch reveals itself during time evolution of the states by the Hamiltonian (\ref{scaledHamil}). Given the initial state in one of the two coherent states, the Hamiltonian (\ref{scaledHamil}) evolves it to their superposition, i.e. $\mib $ and $\posb  $ mainly because time evolution based on the Hamiltonian (\ref{scaledHamil})  does not change central position $\beta$ of these states. The projection of time-evolved state on $\mpb $ is however, imperfect in the sense that there is a possibility for slight deformations that makes the projections only approximately valid. However, the imperfection is periodic so that after times the very coherent states are reproduced. 

\begin{figure*}[t]
\centering
\begin{tabular}{ccc}
	\begin{subfigure}[t]{0.32\textwidth}
		\includegraphics[width= \textwidth]{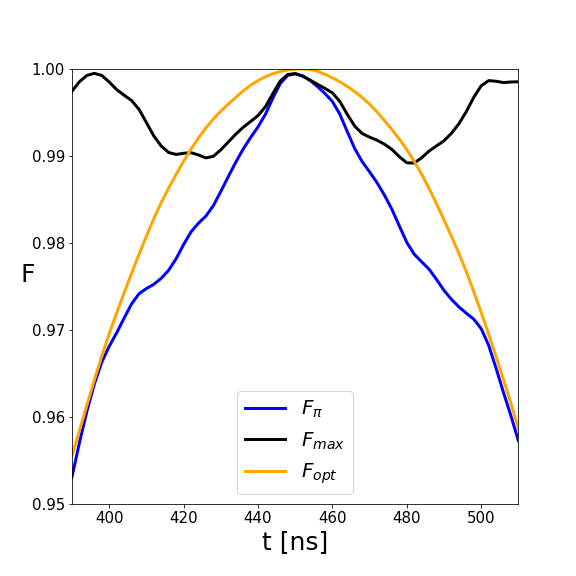}
        \vspace{-2em}\caption*{(a)}
	\end{subfigure} 
	% \hill
	\begin{subfigure}[t]{0.32\textwidth}
				\includegraphics[width= \textwidth]{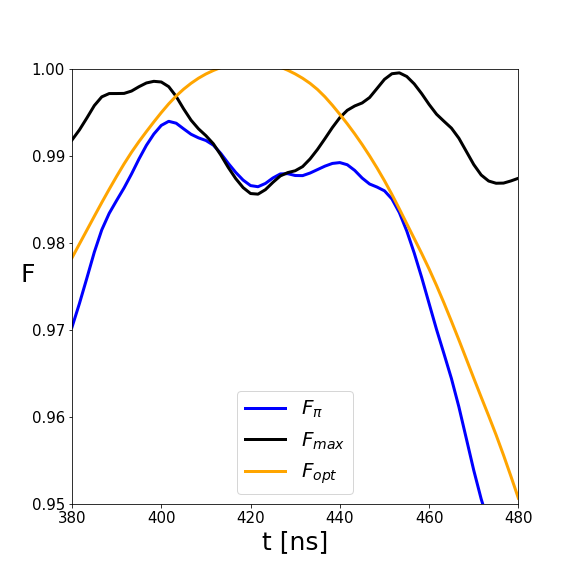} 
    \vspace{-2em}\caption*{(b)}
	\end{subfigure}%

		\begin{subfigure}[t]{0.32\textwidth}
		\includegraphics[width=\textwidth]{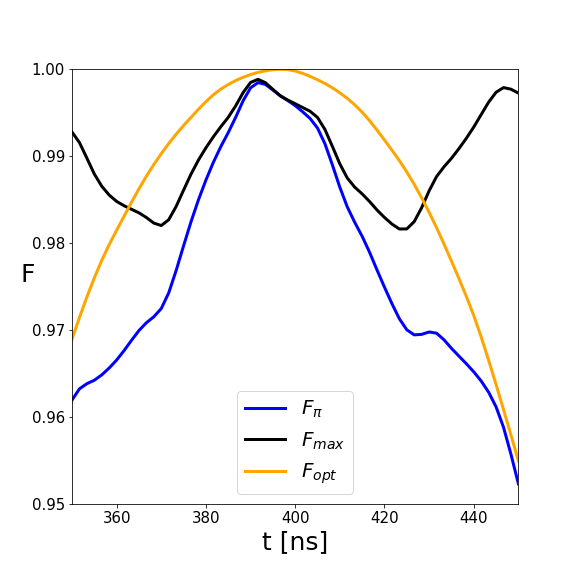}
  \vspace{-2em}\caption*{(c)}
	\end{subfigure}%
\end{tabular}
\caption{\flushboth Fidelity for the X-rotation caused by detuning. All figures (a)-(c) use $K=6.7\,\text{MHz}$ and $\alpha=1$. The fidelity of an $R_X(\pi)$ gate over time shown in blue for $R=0.4$ (a), $R=0.45$ (b) and $R=0.5$ (c). It is bounded by $P_{cs}$ in black and has an idealized version in orange, where the fidelity is scaled by the maximal fidelity at each point. Only the idealized fidelity is a smooth rotation, whereas the actual fidelity shows a sharpened, (a) and (c), or flattened (b) peak, which is caused by the deformation of our states.}\label{fig.f}
\end{figure*}

Let us consider the initial state $| \psi(t=0)\rangle$ being the coherent state $\mib $. This state after time $t_1$ may evolve into $|\psi(t_1)\rangle$, which is not necessarily perfectly projectable on to the coherent states $\mib $. However, mathematically one can find out that the revival time to return on a superposition of the coherent state might be $t_2$, i.e.  
$|\psi(t_2)\rangle  = a \mib   + b \posb $. In order to define a controllable $X$-gate all we need is to make sure that $t_1 = t_2$, i.e. during the revival time the desired rotation takes place to make a rotated state in the space of two coherent states.  Therefore $X$-rotation is erroneous only before the states return to computational subspace. Interestingly, in the rest of this section we show that system parameters can be adjusted so that the $X$-rotation is synchronized with the time period of the deformation.\\

By initializing from $\posb$ and evolving under the Hamiltonian (\ref{scaledHamil}), we can look at the population of finding the state staying in the computational subset by calculating the probability of its projection on $\mib$ and $\posb $ states. So far in literature \cite{Puri2017} this gate has been tested within weak $R$ regime ($R\leq 0.1$) and this show population projection on $\posb $ and $\mib$ follow $\sin^2$ and $\cos^2$ functions, which indicates an ordinary albeit slow rotation on the Bloch sphere, as well as for higher R values focusing on when it becomes an even integer where the eigenspectrum reveals multiple degeneracies \cite{ruiz2023two}.

\begin{figure*}[t]
	\begin{subfigure}[t]{0.65\textwidth}
		\caption*{(a)}\vspace*{-4em}
		\includegraphics[width=1.05\textwidth]{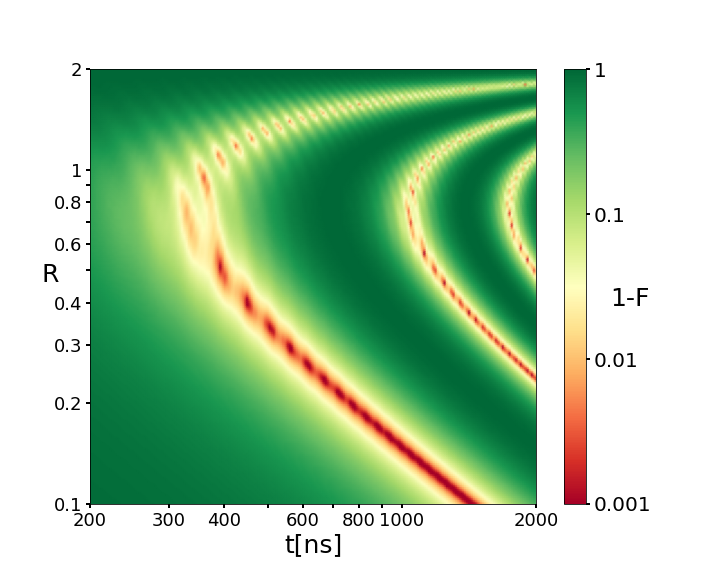}
	\end{subfigure} \hspace*{-3em}
	% \hill
	\begin{subfigure}[t]{0.31\textwidth}
		\caption*{(b)}\vspace*{-2em}
		\includegraphics[width=1.1\textwidth]{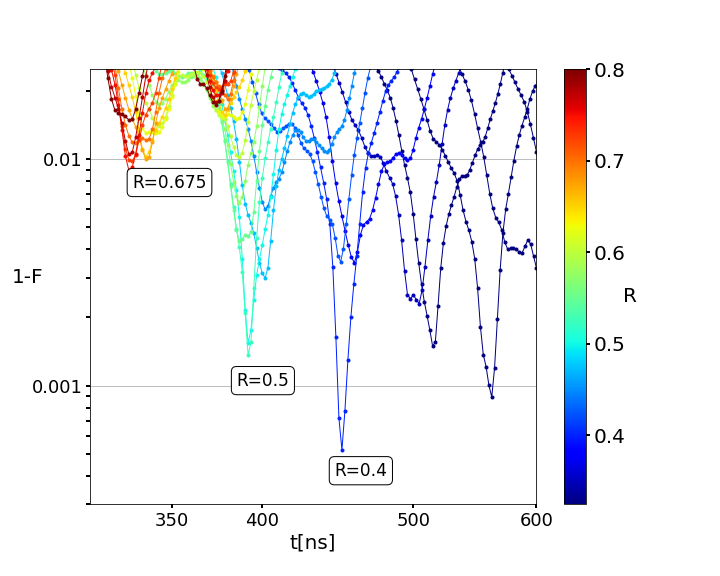} \\ 
	%\end{subfigure}%
	\vspace*{-1em}
	%	\begin{subfigure}[t]{0.24\textwidth}
		\caption*{(c)}\vspace*{-2em}
		\includegraphics[width=1.1\textwidth]{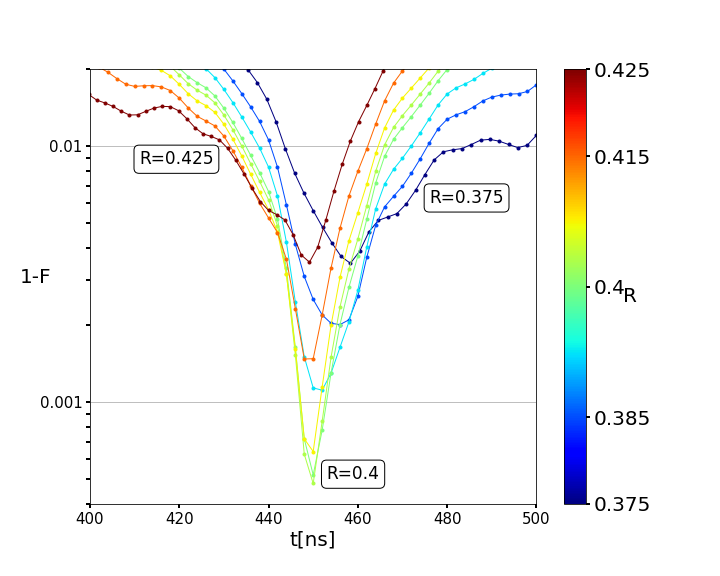}
	\end{subfigure}%
	%\hfill
	\caption{\flushboth Fidelity for the X-rotation caused by detuning. All figures use $K=6.7\,\text{MHz}$ and $\alpha=1$. (a), Error of an $R_X(\pi)$ gate for various values of $R$ and $t$. The red and yellow areas with small errors show the states closest to the expected ones. Fastest gates are realized at $R\approx0.8$ but the line of high fidelity states is disrupted by waves that intersect it. This breaks the line into dots where only specific detunings allow for errors of $<0.001$. (b), Errors for the same gate as in (a). Each line connecting dots uses one $R$ value. The focus is here on $R$ values larger than $0.4$. The peaks at $0.4$ and $0.5$ that are also shown in (d) and (f) are one order of magnitude smaller than the area between them. Since they are sharpened by the detuning a precise control of the gate duration is important, as deviations will lead to a large loss in fidelity. (c), The $R=0.4$ peak from (b) magnified. Large differences in $R$ around $\Delta R =0.025$ diminish the fidelity considerably but smaller variations in the detuning $\Delta \delta=50\, \text{kHz}$ only cause minimally increased errors.}\label{fig2}
\end{figure*}

For larger $R$ the peak of population gets deformed from the trigonometric functions. We find two patterns for the deformation:  the population peaks are sharpened at certain values of $R$, otherwise it is flat and smeared over a larger value, see Fig.\ref{fig2}(d-f).

 Considering that $\mpb $ represent computational states, leakage is defined as projection out of the space. Initiating a qubit at the state $\posb $, one can define the probability for staying in the computational space $\pcs$ at time $t$ as
\begin{equation}
\pcs(t)=|\langle-\beta |e^{-iHt}|\beta\rangle|^2+|\langle\beta|e^{-iHt}|\beta\rangle|^2-e^{-2|\beta|^2}\label{eq. P}
\end{equation}
in which the last term denotes the non-zero overlap between $\mpb$ as they are non-orthogonal states. Ideal rotation is expected to conserve the state within the computational space, i.e. $\pcs(t)=1$.   

Any deviation from ideal case means that the state is partially mapped out of the computational space and therefore $\pcs(t)<1$. Therefore $\pcs(t)$ serves as the upper bound for the fidelity of $X$-rotation. This upper bound however ignores accuracy in targetting specific angle for rotation within the computational subspace.  The black curves in Fig.\ref{fig.f}(a-c) shows the evolution of $\posb $ state in time $t$ for which we evaluate $\pcs(t)$ at each time and show its value in black lines. These values represent  the upper bound of fidelity and one can see their more or less nearly-unity values all almost times $t$. However interestingly the loss in time is neither constant nor increasing, instead it is somewhat periodic . Comparing plots of different detuning $R$ in Fig.\ref{fig.f}(a-c) shows  increasing detuning does not necessarily increase leakage out of the computational space. 

The interesting lesson from studying leakage is that larger detuning does not necessarily result in lower fidelity. In fact one can see that the maximum of fidelity can still reach values close to $1$. For $R=0.4$ and $R=0.5$ one can check that the maximum of $\pcs$ is reached at the time nearly simultaneous to the time required for $\pi/2$ rotation along $X$-axis.  Therefore for these two $R$ values we can perform high fidelity gates, since the evolved state is mapped back into our basis as soon as the gate applies the proper angle of rotation on initial state. Interestingly in the midway between the two values, i.e. $R=0.45$, the loss in fidelity turns out to be large at the time required for $\pi/2$ rotation.

What seems to be a possible strategy is to seek for a high fidelity and fast $X$ gate in systems with large $R$ values without overshooting the minimum of $\pcs$ in the required gate time. Increasing $R$ cannot be arbitrary chosen because there are certain $R$ values at which the gate performs slower, since the required time for rotation may coincide with the low value of $\pcs$.  Fig.\ref{fig2}(a) denotes the leakage upper bound quantity  $1-\pcs(t)$ at different times for different detuning $R$.  Red areas indicates the narrow domain of parameters for high fidelity $X$ rotation performance, with the lowest leakage rate out of code space. The Green area are erroneous parameters that if a dCat qubit is tuned on them they perform erroneous rotation.    Fig.\ref{fig2}(a) shows that the $X$ gate of $R>0.8$ is much slower in comparison to the gate time at  $\lesssim0.8$. 

In Appendix (A) we show that $X$ gate speed $1/K t$ against $R$ using numerical simulation, where $1/t$ goes to zero at the positive even number $R=2,4,6$, see Fig. (\ref{gate_time_X}). This shows the oscillatory behavior of the period.  One also can see that fastest rotation take place at larger $1/Kt$. The maxima of $1/Kt$ depend on the coherent parameter $\alpha$, therefore on detuning. Moreover the minimum gate time if occurs at smaller $R$ will result in higher fidelity gates.  This corresponds to the lower red spots in Fig.\ref{fig2}(a).  Since Fig. (\ref{gate_time_X}) has been worked out for $K=6.7\, \text{MHz}$ and $\alpha=1$, the fastest fast gate takes place at the maximum point of $R\approx0.675$, however according to Fig.\ref{fig2}(b) the error of this point is worse by an order of magnitude compared to the case of $R=0.4, 0.5$. Therefore although the gate at $R=0.675$ is $60 \, \text{ns}$ and serves to be faster than at $R=0.5$, however the speed cannot overcome priority over high fidelity. Therefore the optimization of both parameters are required.

As mentioned earlier, the probability of no leakage in the computational subspace is determined by $\pcs$. Given that the initial state is for instance $\Psi(0)=\posb$, if we normalize the probability of finding the state $\Psi(t)$ by $\pcs$,  the normalized probability interestingly shows perfect trigonometric oscillation between $\posb$ and $\mib$, which is depicted in Fig.\ref{fig2}(d-f) in orange line and gives a sine wave with a smooth peak. The smoothness of the normalized probability indicates that there is no deformation or loss of fidelity once the state stays inside the computational subspace, in other words, the state $\posb$ is not mistaken with $\mib$ at any time. This gives us reason to believe that our understanding of the upper limit for the fidelity is correct and the deformation in the gate performance fidelity is only a limiting factor for reaching low errors of $\lesssim0.001$. Note that the ideal points are unlikely to be achievable in experiments, since the effect of the deformation can not be disabled without disabling the rotation itself.

We therefore suggest that one can make use of careful parameter tuning, so that the period of the deformation matches the gate time. This allows one to use larger detunings and thus faster gate times without sacrificing fidelity.

\section{Z gate for dCats}

The Hamiltonian of a Cat qubit in absence of detuning is $\frac{1}{K}H_{\textup{dCat}} =-\left(\ahd \ahd -\alpha^2\right)\left(\ah\ah -\alpha^2\right)$. This highlights its eigenstates at $\mpa $, since the $\ah\ah$ operator returns an $\alpha^2$ for both states. Rotation along $Z$ axis can be introduced by applying a single photon drive on the resonator, $H_Z=\epsilon_Z \ahd +\epsilon_Z^* \ah $. In the absence of this drive, the eigenstates are the degenerate coherent states $\mpa$, however in the drive presence the degeneracy between $\mpa $ states is lifted and oscillation between them takes place at the Rabi frequency $\Omega_Z=\text{Re}(4\alpha\epsilon_Z)$,  \citep{Grimm2020}. 

For dCat qubits with detuning energy $H_X=\delta \ahd \ah$, we can consider that the total additional terms under the single photon drive is 
\begin{equation}
\frac{1}{K}(H_Z+H_X)=R\left(\ahd +\lambda\right)\left(\ah+\lambda\right)-R\lambda^2,\label{productform_XZ}
\end{equation}
where $\lambda=\epsilon_Z/\delta$ is a complex valued parameter. Here $\lambda$ has a similar role to $\alpha$ in the Cat Hamiltonian as it is defined by the ratio of two parameters and describes an eigenstate of the system. The state $|-\lambda\rangle$ acting on (\ref{productform_XZ}) gives a zero for the $(\ah+\lambda)$ term and is thus an eigenstates with energy $-R\lambda^2$. This means that one of the eigenstates of $H_Z+H_X$ is a perfect coherent state, whose amplitude can be changed by tuning the amplitude of the single photon drive or the detuning of the squeezing drive. The fact that we have a coherent eigenstate is non-trivial, since both $H_Z$ and $H_X$ have no coherent eigenstate, apart from $|0\rangle$ in the $H_X$ case. This coherent eigenstate is the advantage of applying both a single photon drive and a detuning at the same time. On their own no coherent eigenstates are possible and any initialization into coherent states will accumulate some infidelities over time. These can be greatly reduced if the detuning is chosen carefully as demonstrated in the previous chapter, however a perfect eigenstate simplifies calculations and promises higher fidelities.

\begin{figure*}
\centering
\hspace*{-1cm}
\begin{tabular}{cc}

\begin{tabular}{cc}
\setcounter{subfigure}{0}
	\begin{subfigure}[t]{0.28\textwidth}
		\caption{}\vspace*{-1.5em}
		\includegraphics[width=1.1\textwidth]{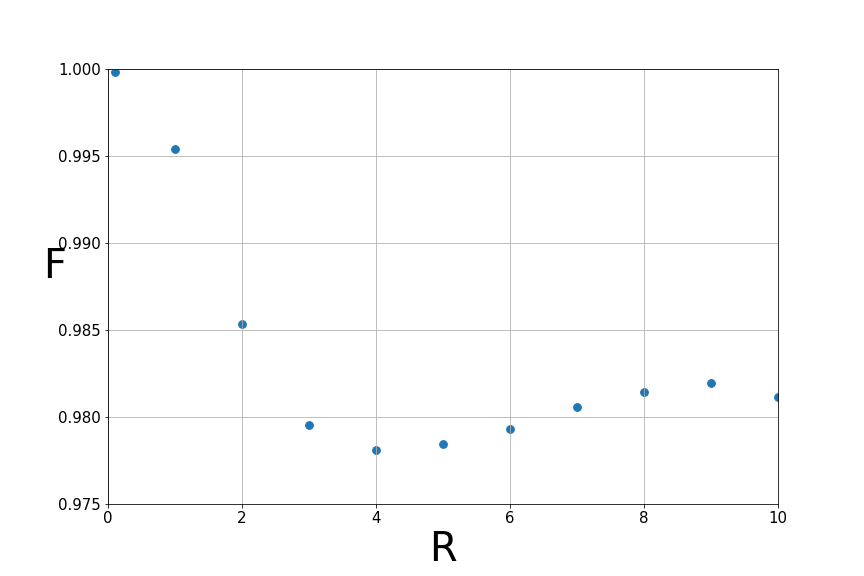}
  \label{a}
	\end{subfigure}&
 \setcounter{subfigure}{1}
        \hspace{-1em}
	\begin{subfigure}[t]{0.28\textwidth}\label{b}
		\caption{}\vspace*{-1.5em}
		\includegraphics[width=1.1\textwidth]{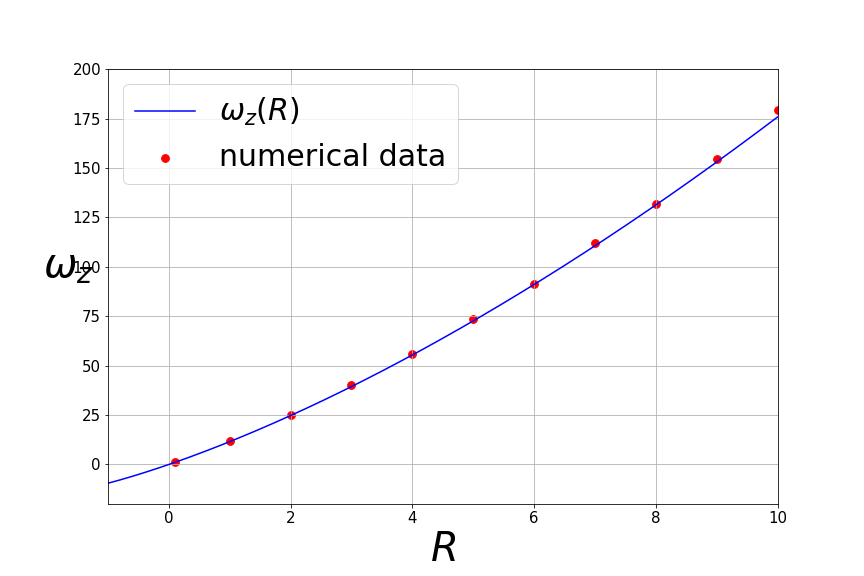}
	\end{subfigure}

 \setcounter{subfigure}{2}
 \hspace{-1em}
 \begin{subfigure}[t]{0.28\textwidth}\label{c}
		\caption{}\vspace*{-1.5em}
		\includegraphics[width=1.0\textwidth]{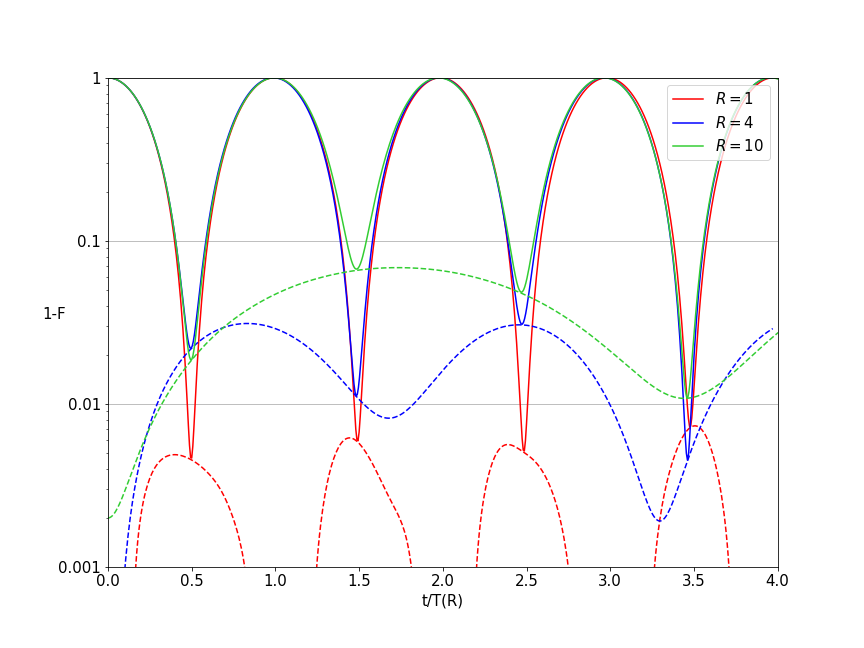}	
	\end{subfigure}

\end{tabular}
\hspace{-2em}
\begin{tabular}{c}
      \setcounter{subfigure}{3}
 	\begin{subfigure}[t]{0.2\textwidth}\label{d}
		\vspace*{1em} \caption{}\vspace*{-1.3em}
		\includegraphics[width=1.1\textwidth]{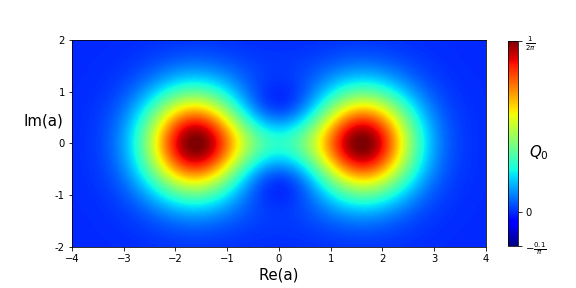}
	\end{subfigure}  \\
  \setcounter{subfigure}{4}
        \begin{subfigure}[t]{0.2\textwidth}\label{e}
		\vspace*{-1.7em} \caption{}\vspace*{-1.3em}
		\includegraphics[width=1.1\textwidth]{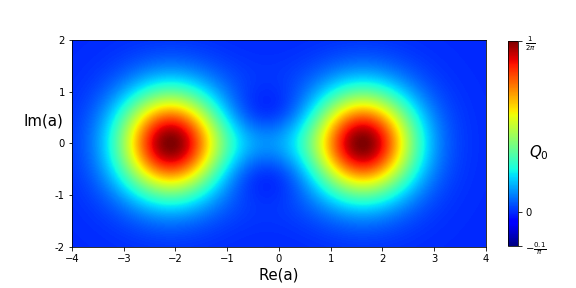}
    \end{subfigure}
\end{tabular}
\end{tabular}
\caption{\flushboth Fidelity and speed of the Z-rotation caused by the single photon drive and modified by detuning. $K=6.7\,\text{MHz}$ and $\alpha=1.63$ are used while $\epsilon_Z$ is given by $\alpha=-\lambda=-\epsilon_Z K /R$. $\epsilon_Z$ therefore scales linearly with $R$ in all subfigures. (a) Fidelity for an $R_Z(\pi)$ gate for varying $R$ values (and $\epsilon_Z$ with it) For $R$ values less than 4 the fidelity drops but then reaches a minimum after which it rises once again. (b), Comparison of the analytical $\omega_Z(R)$ from (\ref{omegaz}) with the numerical values for the frequency of the Z-rotation. They match for both small and large $R$ values. Extremely large values for $R$ correspond to extremely fast gates. (c), Error of the Z-rotation over time for three different $R$ values. The time is normalized for each $R$ value separately so that their points of minimal error align. The full lines indicate the actual error rate of the rotation while the dashed lines show the minimal error rate possible defined in the main text. The errors for $R=1$ reach below $0.01$ consistently over multiple rotations, while for larger values the errors get larger or smaller depending on if the deformation of our states is maximal or minimal. (d),  Husimi $Q$-function $Q_0$ of the states $(|\alpha\rangle+\mia     )/\sqrt{2}$. (e),  Husimi $Q$-function $Q_0$ of the states $(|\alpha\rangle+|\gamma_1\rangle)/\sqrt{2}$ for $R=2$ and the fixed $\epsilon_Z$ value with $\alpha=-\lambda$. The right side of the state can be kept in place and is a perfect eigenstate, while the left state is displaced further to the left to $\gamma_1$, which is not an eigenstate.}\label{fig3}
\end{figure*}

Since the terms $H_Z$ and $H_X$ only show up in addition to $H_{\textup{dCat}} $, we are primarily interested in all three terms at once and in general $\mpa $ are not eigenstates of (\ref{productform_XZ}) and $|-\lambda\rangle$ is not an eigenstate of $H_{\textup{dCat}} $, but since we can change $\alpha$ and $\lambda$ independently from each other, we can make the eigenstates $|\alpha\rangle$ and $|-\lambda\rangle$ coincide by setting $\alpha=-\lambda$.  This means that the state $|\alpha=-\lambda\rangle$ is a coherent state and an eigenstate of the total Hamiltonian \mbox{$H_{\textup{dCat}} +H_Z+H_X$}. This implies that even for extremely large detunings or single photon drive amplitudes a system initialized to $|\alpha\rangle$ stays in that state forever as long as the ratio of $\delta$ and $\epsilon_Z$ is fixed to $-\epsilon_Z/\delta=-\lambda=\alpha$.
However, some problems arise from the state \mbox{$\mia     $}. It is an eigenstate of the dCat  Hamiltonian, but not of the total Hamiltonian. Since $H_Z+H_X$ has only one coherent eigenstate and it already coincides with $|\alpha\rangle$, there is no second eigenstate to match $\mia     $. Similarly to how we defined the coherent states $\mpb $ earlier, we can now define a coherent state $|\gamma_1\rangle$ in the vicinity of $\mia     $ that is not an eigenstate of the total Hamiltonian, but resides at the bottom of the energy well of the total Hamiltonian. Previously this point in the potential was described by $-\alpha$ or $-\beta$, but since we now also applied a single photon drive, it is shifted again. It is located at 
\begin{equation}
\gamma_1=-\frac{\alpha}{2}-\sqrt{\frac{\alpha^2}{4}+\frac{R}{2}}
\end{equation}
and while $\gamma_1$ does not directly depend on $\epsilon_Z$, since $-\epsilon_Z/\delta=\alpha$ is fixed, it indirectly does depend on the amplitude of the single photon drive. In Fig.\ref{fig3}(e) we can see the Husimi $Q$ function of the state $|\gamma_1\rangle+|\alpha\rangle$ which shows the left state slightly shifted to the left when compared to the state $\mia     +|\alpha\rangle$ in Fig.\ref{fig3}(d).

One can also consider the more general case of $\alpha\neq-\lambda$, in which case the positions of the optimal coherent states is given by $0=\gamma^3-\beta^2\gamma-\epsilon_Z/2K$. In this case both states would be affected by deformations and we would acquire fidelity losses in both states. We want to focus on the case of $\alpha=-\lambda$ as this allows us to keep one eigenstate as a coherent state, which should be used as our $|+Z\rangle$ state. This allows us to keep the detuning and single photon drive applied, even if our qubit is supposed to be idling.

Previously the Rabi oscillation in the case of no detuning was described by $\Omega_Z=\text{Re}(4\alpha\epsilon_Z)$. For the case of a finite detuning and $\alpha=-\lambda$ we find the equation
\begin{equation}
\omega_Z/K=-(\gamma_1^2-\alpha^2)^2+R(\gamma_1-\alpha)^2\label{omegaz}
\end{equation}
for the speed of the Z rotation. The equation for $\omega_Z$ can be compared to numerical data in Fig.\ref{fig3}(b) between $R=0.1$ and $R=10$. Since we fixed the ratio of $\epsilon_Z$ and the detuning to be equal to $\alpha$, we change $\epsilon_Z$ alongside $R$. This means that large (or small) $R$ values also coincide with large (or small) $\epsilon_Z$ values. Although the $R=10$ case uses a very large detuning and single photon drive strength and is not particularly useful experimentally, the additional order of magnitude in $R$ shows that the $R_Z(\pi)$ gate can in theory be faster than $0.5\,\text{ns}$ at fidelities over $0.98$, which is visible in Fig.\ref{fig3}(a). In an experimental setup slightly slower gates at higher fidelities should be desirable, since gates at $R=1$ still only take $6.4\,\text{ns}$ but reach fidelities of $>0.995$. An $R$ value in between might sound promising to get fast gates with high fidelities, but Fig.\ref{fig3}(a) shows a minimum of fidelity for $R\approx4$ with no gate duration advantage over higher $R$ values. Therefore one should either choose smaller $R$ values for higher fidelities or higher $R$ values if extremely fast gates are required. The reason for this minimum in fidelity can be understood by looking at Fig.\ref{fig3}(c). The full lines show the error rates for different $R$ values over time, which is normalized so that the periods for different $R$ values match. The dashed lines indicate their lowest error rate possible similar to definition (\ref{eq. P}) in the previous chapter.
Both lines show an oscillation with different periods. For $R=1$ the periods match up, but the overall low errors ensure high fidelity. For $R=4$ the errors increase, but their oscillation slows down when compared to the rotation on the Bloch sphere. However, we still hit only slightly to the left of the first maximum of the errors. For $R=10$ the rotation itself is much faster than the error-oscillation and we can finish the gate before the error reaches its maximum, which increases fidelity slightly when compared to $R=4$. 

Therefore a careful choice of detuning and single photon drive strength is important for the $Z$-rotation. However, when compared to the $X$-rotation it is much faster with similar fidelities. This means that the $X$-rotation can be considered the bottleneck of this gate set and optimizing it should take priority over the $Z$-rotation.

\section{Controlled-NOT Gate for Cats}

After studying the performance of  one-qubit gates, let us now introduce how to couple two Cat qubits by a Control-NOT (CNOT) gate.  We will show that CNOT can be performed with significant improvement in the overall fidelity. 

Consider two Cat qubits, each stabilized in a two-photon driven Kerr nonlinear oscillator, one labelled as `control' and the other one as `target' qubit, with corresponding coherent parameters $\alpha_c$ and $\alpha_t$. The encoding states can be denoted as $\left( a_0\mia_c + a_1\posa_c\right) \otimes \left(b_0\mia_t + b_1\posa_t\right)$.   The most impactful noise on cat qubits is of the form $\hat{O}= f(\alpha) \hat{Z}$ with $f(\alpha)$ being a polynomial function of the coherent parameter.  An implementation of the CNOT has been proposed \cite{Puri2020} where if this error occurs on the control qubit, time evolution of the two-qubit state under CNOT operation preserves the noise on the control qubit and does not spread it to the target qubit. If this error takes place on the target qubit, however, it manifests itself as an error in the control qubit and maintains its $Z$ nature, which indicates that the CNOT on cat qubits protects the noise bias \cite{rahulthesis}.

The performance of CNOT gate on two Cat qubits can be written as interactions between them. The static form of individual qubits have the following typical Hamiltonian: $- K(\hat{a}^{\dagger 2} - \beta^2 ) (\hat{a}^{ 2} - \beta^2 )$. Therefore in the case that the control qubit is at $|0\rangle \equiv |+\beta_c\rangle $, the effective Hamiltonian is $\hat{H}_{|0\rangle} = - K(\hat{a}_c^{\dagger 2} - \beta_c^2 ) (\hat{a}_c^{ 2} - \beta_c^2 ) - K(\hat{a}_t^{\dagger 2} - \beta_t^2 ) ( \hat{a}_t^{ 2} - \beta_t^2 )$,i.e, the Hamiltonian keeps both qubits static when control qubit is $|0\rangle$.

\begin{figure*}[t]
\centering
\begin{tabular}{cc}
	\begin{subfigure}[t]{0.45\textwidth}
		\caption{}\vspace*{-1em}
		\includegraphics[width=\textwidth]{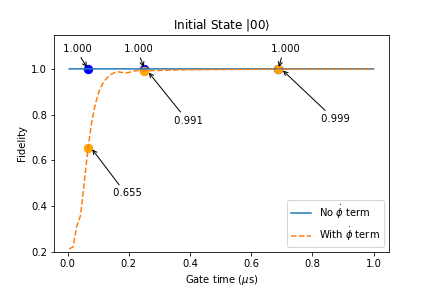}
	\end{subfigure}&
	\begin{subfigure}[t]{0.45\textwidth}
		\caption{}\vspace*{-1em}
		\includegraphics[width=\textwidth]{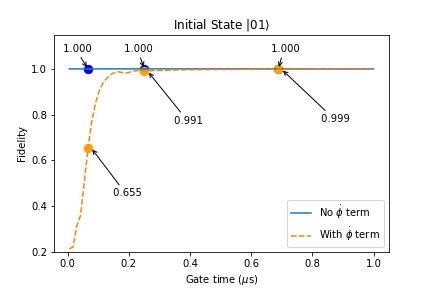}
	\end{subfigure}\\
	\multirow{2}{*}{
	\begin{subfigure}[t]{0.45\textwidth}
		\caption{}\vspace*{-1em}
		\includegraphics[width=\textwidth]{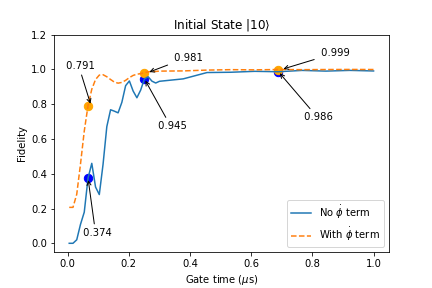}	
	\end{subfigure}}&
	\begin{subfigure}[t]{0.45\textwidth}
		\caption{}\vspace*{-1em}
		\includegraphics[width=\textwidth]{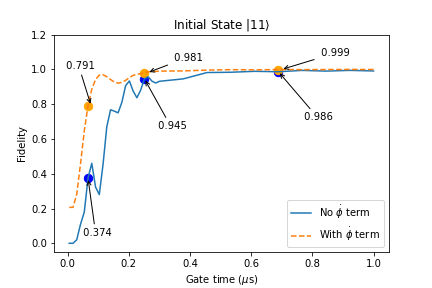}
	\end{subfigure}
\end{tabular}
\caption{\flushboth Fidelity of CNOT for each of the basis states defined as overlap of numerically calculated state with expected state. The corresponding fidelity values are shown for both $\Dot{\phi}$ and $\Dot{\phi}$-free Hamiltonian for gate time $T=0.0665, 0.251$ and $0.689~\mu$s. (a), (b) show the case when control qubit is at $|0\ket$ and $\Dot{\phi}$-free Hamiltonian has fidelity of 1 for all gate times, while $\Dot{\phi}$-Hamiltonian performs worse for fast gates. (c), (d) demonstrate the case when control qubit is at $|1\ket$. Here, the $\Dot{\phi}$-Hamiltonian performs much better especially for fast gates.}\label{lin_fidel}
\end{figure*}

However, when control qubit is at $|1\rangle \equiv |-\beta_c\rangle$ a different effective Hamiltonian is expected as the result of applying CNOT. In this case, the control part is still static Hamiltonian, however it has been shown that CNOT gate evolves the state of target qubit $|\pm \beta_t\rangle$ by introducing an additional $\phi$ freedom in the target qubit essentially replacing $\beta_t $ by $\beta_t e^{i\phi(t)}$. By changing $\phi(t)$ from $0$ to $\pi$ adiabatically, we rotate the target qubit and execute a NOT gate on it, conditioned on when control qubit is at $|1\rangle$. It should be noted that this rotation takes the target qubit out of the Bloch sphere except at the beginning and the end of the CNOT gate.  One can show that when the control qubit is at $|1\rangle$, the expected Hamiltonian from CNOT is: $\hat{H}_{|1\rangle} =  K(\hat{a}_c^{\dagger 2} - \beta_c^2 ) (\hat{a}_c^{ 2} - \beta_c^2 ) - K(\hat{a}_t^{\dagger 2} - \beta_t^2 e^{-2i \phi(t)} ) ( \hat{a}_t^{ 2} - \beta_t^2 e^{2i \phi(t)}) - \dot{\phi}(t) \hat{a}_t^\dagger \hat{a}_t$ in which the last term serves as a way to remove an additional geometric phase \cite{Puri2020,rahulthesis}.

Combining these two conditional Hamiltonians into one, the two interacting Cat qubits under CNOT gate takes the following form:  \cite{Puri2020} 
\begin{align}\label{CNOT}
    \hat{H} = &- K(\hat{a}_c^{\dagger 2} - \beta_c^2 ) (\hat{a}_c^{ 2} - \beta_c^2 ) \nonumber\\
    &-K\Big[ \hat{a}_t^{\dagger 2} - \beta_t^2 e^{-2i\phi(t)}\Bigg(\frac{\beta_c - \hat{a}_c^\dagger}{2\beta_c}\Bigg) -\beta_t^2 \Bigg(\frac{\beta_c +\hat{a}_c^\dagger}{2\beta_c}\Bigg) \Big]\nonumber\\
    & \times\Big[ \hat{a}_t^{2} - \beta_t^2 e^{2i\phi(t)}\Bigg(\frac{\beta_c - \hat{a}_c}{2\beta_c}\Bigg) -\beta_t^2 \Bigg(\frac{\beta_c +\hat{a}_c}{2\beta_c}\Bigg) \Big]\nonumber\\
    &-\Dot{\phi}(t) \frac{\hat{a}_t^\dagger\hat{a}_t}{4\beta_c} (2\beta_c - \hat{a}_c^\dagger - \hat{a}_c). 
\end{align}

After the gate is executed, the geometric phases difference between $|\pm\beta_t\rangle$ turns out to be $\Phi^- - \Phi^+ = 4 \phi(t) \beta_t^2 e^{-2 \beta_t^2}/(1-e^{-4 \beta_t^2})$. This indicates that for large coherence parameter $\beta_t$ the geometric phases are nearly equal. This additional geometric phase $\Phi = \Phi^+ = \Phi^-$ is corrected by the $\dot{\phi}(t)$ term in Eq.\ref{CNOT} .  The phase $\phi(t)$ is supposed to adiabatically vary in time, therefore $\dot{\phi}(t)$ should be much smaller than the energy barrier between the two states.

The last term in Eq.~\ref{CNOT} (referred as the $\Dot{\phi}(t)$ term) is similar to an approximate detuning term that we studied for single qubit gates, despite the fact that their origins are quite different. Although introduced as a method for correcting relative Z error, it is much more impactful towards improving the fidelity as we will see next.

Given that target qubit is prone to a relative $Z$ error, it has been proposed in \cite{Puri2020} that the last term of Hamiltonian (Eq.\ref{CNOT}) can help to correct the $Z$ error. Here we will see another application of the term for improving the overall fidelity of the gate. We will explore how the term impacts the evolution of the basis states and eventually attempt to better understand the origin of the fidelity improvements. 

First let us discuss the nature of the last term interaction between the two qubits. Given that Hamiltonian (Eq.\ref{CNOT})  is time depending, let us consider its instantaneous eigenstates and eigenvalues to be $|\psi_n(t)\rangle$ and $E_n(t)$,  following the relation: $\Hat{H}(t)|\psi_n(t)\rangle = E_n(t)|\psi_n(t)\rangle $. The adiabatic theorem \cite{sakurai1967advanced} states that the system in one of its eigenstate $|\psi_n(t)\rangle$ remains in the instantaneous eigenstate as the Hamiltonian is continuously changed with time, provided that the corresponding eigenvalue is sufficiently gapped from the rest of eigenvalues and the rate of change of the Hamiltonian is slow enough compared to the minimum energy gap between the state of other eigenstates.

The concept of Shortcut to Adiabaticity (STA) \cite{berry2009transitionless} has been introduced to speed up the adiabatic evolution by introducing corrections to the original time dependant Hamiltonian.  A reverse engineering approach can be used to get to the correction term. Our aim of maintaining the state on a single eigenstate means that we want to obtain the state during evolution to remain as $\exp(-i\int_0^t dt' E_n(t'))\ |\psi_n(t)\rangle$, namely $|\Psi_n(t)\rangle$. Projecting the state at each time on the initial state $\langle \Psi_n(0)|$ and summing over all states, makes the following unitary operator $\hat{U}(t)=\sum_n |\Psi_n(t)\rangle \langle \Psi_n(0)|$. This definition helps to define $i \partial_t \hat{U}(t) = \hat{\mathcal{H}}\hat{U}(t)$, with  $\hat{\mathcal{H}(t)}$ being $i [\partial_t \hat{U}(t)] \hat{U}^\dagger(t)$.  By expanding the relation based on expansion of  $|\Psi_n(t)\rangle$ state in terms of $|\psi(t)\rangle$ states one can find that $\hat{\mathcal{H}}(t)= \hat{H}(t) + \hat{h}(t)$, in which $H(t)$ is the original Hamiltonian and $\hat{h}(t) \equiv i\sum_n |\partial_t \psi(t)\rangle\langle \psi_n(t)|$. Therefore the restriction on the rate of change of Hamiltonian can be avoided by introducing the correction term $\hat{h}(t)$ and this helps to drop the condition of adiabaticity.

Although the correction term $\hat{h}(t)$ allows us to drop the adiabatic condition and allowing  gate to act faster, in practice the correction term is difficult to evaluate as it requires the knowledge of instantaneous eigenstates $|\psi_n(t)\rangle$. However, for the special case of the bias-preserving CNOT Gate, the term can be obtained exactly when projected to the computational subspace. During gate execution if control qubit is at $|1\rangle$ ($=\mibc$), the expected  target qubit state will be $|\mspace{-3mu}\pm \mspace{-5mu} \beta_t e^{i\phi(t)}\rangle$ with $\phi(t)$ varying from 0 to $\pi$ monotonically and adiabatically. The effective adiabatic target qubit Hamiltonain without the additional $\dot{\phi}(t)$ term is given as,
\begin{equation}\label{effectiveHamil1}
    \hat{H}(t) = - K(\hat{a}_t^{\dagger 2} - \beta_t^2 e^{-2i \phi(t)} ) ( \hat{a}_t^{ 2} - \beta_t^2 e^{2i \phi(t)}) 
\end{equation}

As a reminder the subindex $t$ indicates target qubits and the variable $(t)$ indicates passing time.  The corresponding unitary operator that acts on the target qubit is given by: 
\begin{eqnarray}
\hat{U}(t)&=&e^{-i \int_0^t dt' E_0^+} \cplusket \cplusbra \nonumber \\ && + e^{-i \int_0^t dt' E_0^-} \cminusket \cminusbra 
\end{eqnarray}
with $\cpmket = (|\mspace{-4mu}+\mspace{-4mu}\beta_t e^{i\phi(t)}\rangle \pm |\mspace{-4mu}-\mspace{-4mu}\beta_t e^{i\phi(t)}\rangle) N^\pm_{\beta_t}$ and normalization factor $N^\pm_{\beta_t}$ 
of the Cat states. Since both states are degenerate under the Cat Hamiltonian, eigenvalues of these states are $E_0^-=E_0^+$.  

The correction term to the Hamiltonian Eq.\ref{effectiveHamil1} is given by  $\hat{h}(t) = i ( \Dcplusket \cplusbra + \Dcminusket \cminusbra)$. By expanding the partial derivative with respect to time, $\Dcpmket = \partial_t  (|\mspace{-6mu}+\mspace{-6mu}\beta_t e^{i\phi(t)}\big{\rangle} \pm |\mspace{-6mu}-\mspace{-6mu}\beta_t e^{i\phi(t)}\big{\rangle})N_{\pm}$. After substituting the definition $\cpmket= \sum_n e^{-\beta^2/2} (\pm \beta e^{i\phi(t)})^n |n\rangle/\sqrt{n!}$ and some lines of algebra, one can show that $\Dcpmket = i \dot{\phi}(t) \ahd \ah \cpmket$. By substituting this into the correction term one finds,
\begin{eqnarray}
    \hat{h}(t)= - \dot{\phi} \ahd \ah && \left(\cplusket \cplusbra \right. \nonumber \\
    && \left. + \cminusket \cminusbra \right)
    \end{eqnarray}

We obtain that the correction term is exactly equal to the $\Dot{\phi}(t)$ term when projected within the instantaneous target qubit subspace.

Surprisingly below we will see that $\dot{\phi}$ term not only improves the gate speed, but also it is instrumental in improving fidelity when control qubit is at $|1\rangle$. To show this let us now consider two different types of evolution, with and without $\dot{\phi}(t)$ term.  For simplicity, we consider $\phi(t)$ linearly changes with time, i.e. $\phi(t) = \pi t/T$ with $T$ being the CNOT gate time. Under such consideration, we use the Hamiltonian (Eq.\ref{CNOT}) to evolve quantum states in the computational basis up to the gate time $T$ for different values of $T$.   Figure (\ref{lin_fidel}) compares the fidelity of the gate in the presence and absence of $\dot{\phi}$ term in the four computational state evolution for different gate time $T$.  Each point in the plot corresponds to a complete CNOT gate, hence a shorter time implies a faster overall gate. The dashed line represents the action of the Hamiltonian (Eq.~\ref{CNOT}) without the $\dot{\phi}(t)$ term, the solid line includes the action of the mentioned term.

The $\dot{\phi}(t)$ case demonstrates a drop in fidelity for fast gates when control qubit is at $|0\rangle_c$. As gate-time increases, the difference in fidelity decreases sharply. However, the situation becomes the opposite when control qubit is at $|1\rangle_c$. The performance is better with $\dot{\phi}(t)$ term, even for larger gate-times. However, the improvement is much more prominent for faster gates ($\lesssim 0.3~\mu$s). The error shown in the plots, arise due to leakage outside code space since we ignore terms which are exponentially decaying in $\alpha_c,\alpha_t$.

An interesting feature of Fig.~\ref{lin_fidel} is the oscillations in fidelity instead of steady growth. To study these points, we evolved the gates for specific fixed values of total gate time, corresponding to some peaks and troughs in the plots. Simultaneously, the overlap with the instantaneous expected state is evaluated as the qubits evolve. Note that this is not an overlap with the final state after gate execution but with respect to the ideally expected state of the qubit at that point of the evolution, which would usually be outside the Bloch sphere. Here we find oscillations in fidelity, although the location of peaks and troughs change with gate time. We noticed that in Fig.~\ref{lin_fidel}, peaks in the fidelity, both for with and without $\dot{\phi}$ Hamiltonian, occur precisely when the gate ends at a peak in the instantaneous fidelity evolution and vice versa for the troughs.

\section{Conclusion}
DCat  qubits provide an interesting modification to cat qubits where instead of trying to get rid of detuning completely, we instead focus on what values of this term still allows us to perform efficient single qubit gates. The detuning introduces deformations. However, for some  values of $R=\delta/K$ ($=0.4$ and $0.5$), we find that the time period of deformation exactly matches the $X$-gate time, allowing for high fidelity gate. In fact, we find a range of (R,T) values where peak in fidelity can be reached, although not all peaks are equal in value. 

Even after introducing a single photon drive for $Z$-rotation, we can find a coherent state eigenstate of the full Hamiltonian. Using that and pushing one of the coherent states to the minima of the Hamiltonian allows us to implement fast $Z$-rotations before error accumulation due to detuning becomes too high.

Finally we explore CNOT gates in usual cat qubits where we notice that an approximate detuning term allows for a high fidelity gate. We show that the improvement here comes from the equivalency of the detuning term to Shortcut to Adiabaticity correction, restricted within the codespace.

\begin{acknowledgments}
    This research received funding from Horizon Europe (HORIZON) Project: 101113946 OpenSuperQPlus100.
\end{acknowledgments}

\appendix
\section{X-Gate time variation for usual cat qubits}
\begin{figure}[h]
\includegraphics[width=1\linewidth]{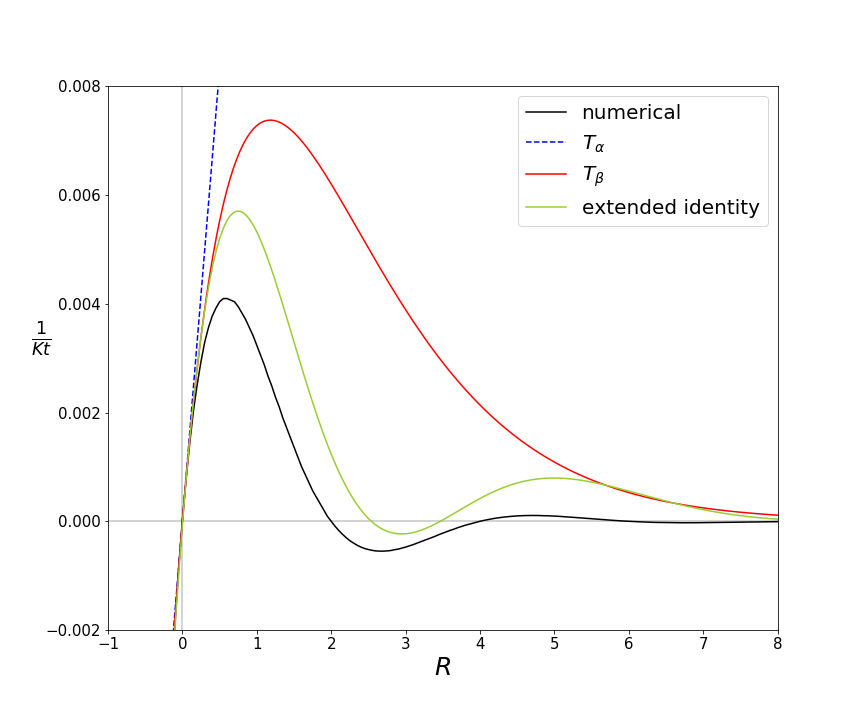}
\caption{\flushboth Gate speed of the $R_X(\pi)$ gate against $R$. $K=6.7\,\text{MHz}$ and $\alpha=1.63$ were used. Negative values for the gate time correspond to a rotation in the opposite direction. The black line shows numerical data, for which half a rotation is measured by stopping at the point where the fidelity of the gate is maximal. The numerical data indicates that $1/t$ goes to zero for $R$ being a positive, even integer. These zero points have been evaluated by interpolating between neighbouring $R$ values. $T_\alpha$ and $T_\beta$ are approximations for small detunings and are therefore tangent to the numerical data around $R=0$. The extended identity approach uses additional coherent states in the basis and tries to cover leakage out of the codespace. It predicts the oscillatory behavior of the period although not accurately. The calculations could be improved by using even more coherent states in the basis}\label{gate_time_X}
\end{figure}

We can study the impact of single qubit detuning term on cat qubits by looking at its corresponding projection to the codespace. Defining the projector on the codespace as $\hat{I} = |\alpha\rangle \langle\alpha| + \mia      \langle-\alpha|$, the projection of the detuning term is
\begin{equation}
    \hat{I} \delta \hat{a}^\dagger \hat{a} \hat{I} = \delta \alpha^2 \hat{I} - \delta\alpha^2 e^{-2\alpha^2} \sigma_x
\end{equation}
where $\sigma_x = |\alpha\rangle \langle-\alpha| + |\alpha\rangle \langle-\alpha| $. Hence, the time period for $R_X(\pi)$ is given by
\begin{equation}
    T = \frac{\pi}{2 \delta\alpha^2 e^{-2\alpha^2}}
\end{equation}
The expression shows that inverse of gate time ($1/T$) should scale linearly with $\delta$. However, performing a numerical evaluation of the gate time (defined as time of maximum overlap with expected state) shows an oscillatory behaviour shown in Fig.\ref{gate_time_X} around $1/T = 0$. These points correspond to diverging of the gate time (i.e, $X$ gate does not get executed at all). An interesting feature of the oscillatory behavior is that these divergences specifically occur when R is an even integer.

The analytic expression is not accurate since we are using non-orthogonal states for constructing the identity operator. As a result, even within the codespace, the identity holds up to an exponentially small error. Also, we have ignored other states in the phase space which can have contribution due to leakage out of codespace. A possible extension of the projector in that case can be
\begin{equation}
    I_{ext} = |C_\alpha^+\rangle \langle C_\alpha^+| + |C_\alpha^-\rangle \langle C_\alpha^-|  + |C_{i\alpha}^{+}\rangle \langle C_{i\alpha}^{+}|  + |C_{i\alpha}^{-}\rangle \langle C_{i\alpha}^{-}| 
\end{equation}
where the states are defined as $|C_\alpha^\pm\rangle =\mathcal{N}_{\alpha}^\pm (|\alpha\rangle \pm \mia      ), |C_{i\alpha}^{\pm}\rangle = \mathcal{N}_{\alpha}^\pm  (|i\alpha\rangle \pm |-i\alpha\rangle )$. It should be noted here that $i$ is a phase of $\alpha$,which is outside the Bloch sphere, and not the relative phase between $|\pm\alpha\ket$, which do belong to the computational subspace. Hence, $|C_{i\alpha}^{\pm}\ket$ states are outside the Bloch sphere of $|C_{\alpha}^{\pm}\ket$ states. Projecting the detuned Hamiltonian using this extended identity gives the following expression for gate time
\begin{align}
    \frac{1}{KT} = \frac{1}{\pi K} \Big( R &(-z + i b_+ b_- w) + \frac{|\alpha|^4}{2} (b_+^2 + b_-^2) +\nonumber\\ &
    \frac{\epsilon_2}{K} (\sqrt{1-b_+^2}   - \sqrt{1 + b_-^2})  \Big)
\end{align}
where
\begin{align*}
    z &= \frac{|\alpha|^2}{2} (\tanh{|\alpha|^2} - \coth{|\alpha|^2})\\
    w &= \frac{|\alpha|^2}{2} (\tanh{|\alpha|^2} + \coth{|\alpha|^2})\\
    b_+ &= \frac{\cos{|\alpha|^2}}{\cosh{|\alpha|^2}} \\
    b_- &= i\frac{\sin{|\alpha|^2}}{\sinh{|\alpha|^2}}
\end{align*}
The expression allows us to recover the oscillatory behavior expected from the numerical analysis, as in Fig.~\ref{gate_time_X}, although the locations of the zeros are incorrect. The expression can be made more accurate by extending it to more coherent states.

The red line in Fig.~\ref{gate_time_X} refers to the smaller identity expression but corresponding to the dCat  qubits i.e, with $\beta = \sqrt{\alpha^2 + R/2}$. Since, now the $\beta$ value also depends on $R$, the final gate time expression is no longer linear.

\section{Divergence of X-gate time}

An interesting feature of Fig.\ref{fig2}(a)  is that fidelity is nearly 0 for all gate times at $R=2$. In Fig.\ref{gate_time_X} the same phenomenon shows up at $R=2$ as infinite gate time of the $X$-gate. As shown in \cite{ruiz2023two}, the detuning term modifies the overall eigenspectrum of the Hamiltonian. In fact, for $R = 2m$, where $m\in\mathbb{N}$, the excited level pairs $|\alpha_{e,i}^\pm\rangle$ become exactly degenerate for $i\leq m$. This phenomenon also causes the X-gate time to diverge.

The calculations in \cite{ruiz2023two} give an analytic expression for $\mpa $ based on the current eigenstates when $R =2 $. Based on that, for usual cat qubits, we find that the overlap with the target state is always exponentially low $|\langle -\alpha| e^{-iHt}|\alpha\rangle|^2 \sim e^{-|2\alpha|^2} $ for all gate times t, and evolution under the detuned hamiltonian does not implement the X-gate. Evaluating the full expression for overlap, with $\alpha = 1.63$ gives a maximum value of $|\langle -\alpha| e^{-iHt}|\alpha\rangle|^2 \lesssim 7 \times 10^{-4}$ while for larger $\alpha$, the overlap is even less. The same analysis can be done for $R = 2m$ in general as well. 

For dCat  qubits, we carry out an approximate analysis. In the limit of large $\alpha$, we can define the shift as
\begin{align}
    \eta &= \beta - \alpha = \sqrt{\alpha^2 + R/2} - \alpha\nonumber\\
    &\approx \frac{R}{4\alpha} = \frac{1}{2\alpha}
\end{align}
Similarly, the dCat  states can be approximately written as follows
\begin{align}
    |\beta\ket &= \hat{D}(\eta) |\alpha\ket \approx \Big(\hat{I} + \eta (\ahd - \ah)\Big) |\alpha\ket\nonumber \\
    &= |\alpha\ket + \delta|\alpha,1\ket\\
     |-\beta\ket &= \hat{D}(-\eta) |-\alpha\ket \approx \Big(\hat{I} - \eta (\ahd - \ah)\Big)|-\alpha\ket\nonumber \\
     &= |\alpha\ket - \delta|\alpha,1\ket
\end{align}

By using the expansion of $\mpa $ as before, can show that the overlap is still exponentially small i.e, $|\langle -\beta| e^{-iHt}|\beta\rangle|^2 \sim e^{-|2\alpha|^2} $. Evaluating the full expression for $\alpha= 1.63$, the maximum overlap is quite low, $|\langle -\beta| e^{-iHt}|\beta\rangle|^2 \lesssim 3 \times 10^{-4}$. 
The above calculation holds in the limit of $R\ll2\alpha$, however it appears that the approximation works well even for $R=2,\alpha=1.63$ as in Fig.~\ref{fig2}.

\end{document}